\documentclass[%twocolumn,showpacs,
preprintnumbers]{revtex4}
\usepackage{graphicx}

\usepackage{epsbox}
\usepackage{amsmath,amssymb}
\usepackage{booktabs}
\usepackage{bm}

\begin{document}
\date{\today}
\preprint{\vtop{
{\hbox{YITP-10-43}\vskip-0pt
 \hbox{KANAZAWA-10-04} \vskip-0pt
%                 \hbox{hep-ph/1006.????} 
}
}
}

\title{Tetra-quark Systems in Heavy Mesons\footnote{Invited 
talk at the workshop on {\it New Frontier of QCD 2010 
(NFQCD 2010)}, Jan. 18 -- Mar. 19, 2010, Yukawa Institute for 
Theoretical Physics, Kyoto University, Kyoto, Japan. 
} 
\\
-- $\bm{D_{s0}^+(2317),\,\,X(3872})$ and related --}
%        %You can use \\ for explicit line-break }

\author{Kunihiko Terasaki} %\email{terasaki@yukawa.kyoto-u.ac.jp}
\affiliation{ Yukawa Institute for Theoretical Physics, 
Kyoto University, Kyoto 606-8502, Japan,\\
Institute for Theoretical Physics, Kanazawa University, 
Kanazawa 920-1192, Japan
}
\numberwithin{equation}{section}
\thispagestyle{empty}

\begin{abstract}
Typical candidates of open- and hidden-charm tetra-quark mesons are 
studied through their decays and productions, and are compared with 
conventional mesons. 
In addition, it is proposed how to confirm experimentally that they are 
tetra-quark mesons. 
\end{abstract}

\maketitle

\section{Introduction}

Tetra-quark mesons can be classified into the following four groups 
in accordance with the difference of symmetry property of their flavor  
wavefunctions (wfs.)~\cite{Jaffe,D_{s0}-KT}, 
%%%%%%%%%%%%%%%%%%%%%%%%%%%%%%%%%%%%%%%%%%%%%%%%%%%%%%%%%%%%%%%%%%%%%%%%
\begin{equation} 
\{qq\bar q\bar q\} =  
[qq][\bar q\bar q] \oplus (qq)(\bar q\bar q)  
\oplus \{[qq](\bar q\bar q)\oplus (qq)[\bar q\bar q]\},\,\,\,  
(q=u,d,s,c),  
                                                   \label{eq:4-quark} 
\end{equation} 
%%%%%%%%%%%%%%%%%%%%%%%%%%%%%%%%%%%%%%%%%%%%%%%%%%%%%%%%%%%%%%%%%%%%%%%%
where parentheses and square brackets denote symmetry and anti-symmetry, 
respectively, of flavor wfs. under exchange of flavors between them. 
Each term on the right-hand-side (r.h.s.) of Eq.~(\ref{eq:4-quark}) 
is again classified into two groups  
%%%%%%%%%%%%%%%%%%%%%%%%%%%%%%%%%%%%%%%%%%%%%%%%%%%%%%%%%%%%%%%%%%%%%%%%
with 
${\bf \bar 3_c}\times{\bf 3_c}$ and ${\bf 6_c}\times {\bf \bar 6_c}$    
%%%%%%%%%%%%%%%%%%%%%%%%%%%%%%%%%%%%%%%%%%%%%%%%%%%%%%%%%%%%%%%%%%%%%%%%
of the color $SU_c(3)$, 
%%%%%%%%%%%%%%%%%%%%%%%%%%%%%%%%%%%%%%%%%%%%%%%%%%%%%%%%%%%%%%%%%%%%%%%%
which can provide colorless tetra-quark states. 
%%%%%%%%%%%%%%%%%%%%%%%%%%%%%%%%%%%%%%%%%%%%%%%%%%%%%%%%%%%%%%%%%%%%%%%%
The force between two quarks~\cite{color} is attractive (or repulsive) 
when they are of ${\bf \bar 3_c}$ (or ${\bf 6_c}$), so that the 
${\bf \bar 3_c}\times{\bf 3_c}$ state is taken as the lower lying one. 
Narrow widths of the open- and hidden-charm tetra-quark mesons with 
${\bf \bar 3_c}\times{\bf 3_c}$ can be understood by a small overlap of 
color and spin wfs. 
On the other hand, the light scalar mesons~\cite{PDG08}, $a_0(980)$, 
$f_0(980)$, $\kappa(800)$ and $\sigma(600)$, in particular, their mass 
hierarchy and the approximate degeneracy between $a_0(980)$ and 
$f_0(980)$ can be easily understood in the $[qq][\bar q\bar q]$ scheme. 
However, in this case, the corresponding small overlap of color and spin 
wfs. is not guaranteed, because QCD is non-perturbative and states with 
${\bf \bar 3_c}\times{\bf 3_c}$ and ${\bf 6_c}\times {\bf \bar 6_c}$ 
can largely mix with each other at such a low energy scale, so that 
they are not necessarily narrow. 
When it is required that the total wfs. of $[qq]$ and $(qq)$ are 
antisymmetric as in the flavor symmetry limit, their spins are 0 and 1, 
respectively, because the color wf. is antisymmetric for %the
${\bf \bar 3_c}$.  
Therefore, the spin and parity of (at least, dominant components of) 
$[qq][\bar q\bar q]$ and $[qq](\bar q\bar q)\pm (qq)[\bar q\bar q]$ 
mesons with ${\bf \bar 3_c}\times{\bf 3_c}$ are $J^P=0^+$ and $1^+$, 
respectively. 
For the same reason, $(qq)(\bar q\bar q)$ can have 
$J^P = 0^+,\,1^+,\,2^+$. 
However, we ignore it, because no candidate of $(K\pi)_{I=3/2}$ state 
which can be given by the $(qq)(\bar q\bar q)$ state has been 
observed~\cite{K-pi-3/2} in the region $\lesssim 1.8$ GeV in contrast 
with the theoretical expectation~\cite{Jaffe}. 
For more details, see Refs.~\cite{KT-Hadron2003,HT-isospin,ECT-talk}. 

%%%%%%%%%%%%%%%%%%%%%%%%%%%%%%%%%%%%%%%%%%%%%%%%%%%%%%%%%%%%%%%%%%%%%%%%
\section{Open-Charm Scalar Mesons}
%%%%%%%%%%%%%%%%%%%%%%%%%%%%%%%%%%%%%%%%%%%%%%%%%%%%%%%%%%%%%%%%%%%%%%%%

$D_{s0}^+(2317)$ was discovered~\cite{Babar-e^+e^-D_{s0},CLEO-D_{s0}} 
through the $D_s^+\pi^0$ channel in inclusive $e^+e^-$ annihilation, 
while  no signal of resonance peak at the same energy in the radiative 
$D_s^{*+}\gamma$ channel has been observed. 
Therefore, a severe constraint 
%%%%%%%%%%%%%%%%%%%%%%%%%%%%%%%%%%%%%%%%%%%%%%%%%%%%%%%%%%%%%%%%%%%%%%%%
\begin{equation}
R(D_{s0}^+(2317))_{\rm CLEO}
= \frac{\Gamma(D_{s0}^+(2317) \rightarrow D_{s}^{*+}\gamma)}
{\Gamma(D_{s0}^+(2317) \rightarrow D_{s}^{+}\pi^0)}\Biggl|_{\rm CLEO}
< 0.059    
                                             \label{eq:ratio-D_{s0}}
\end{equation}
%%%%%%%%%%%%%%%%%%%%%%%%%%%%%%%%%%%%%%%%%%%%%%%%%%%%%%%%%%%%%%%%%%%%%%%%
was given by the CLEO~\cite{CLEO-D_{s0}}. 
In the case of $D_s^{*+}$, the ratio of decay rates has been measured 
as~\cite{PDG08} 
%%%%%%%%%%%%%%%%%%%%%%%%%%%%%%%%%%%%%%%%%%%%%%%%%%%%%%%%%%%%%%%%%%%%%%%%
\begin{equation}
R(D_{s}^{*+})_{\rm exp} 
= \frac{\Gamma(D_{s}^{*+} \rightarrow D_{s}^{+}\pi^0)}
{\Gamma(D_{s}^{*+}\rightarrow D_{s}^{+}\gamma)}\Bigr|_{\rm exp}
= 0.062\pm 0.008. 
                                             \label{eq:ratio-D_s^*-exp}
\end{equation}
%%%%%%%%%%%%%%%%%%%%%%%%%%%%%%%%%%%%%%%%%%%%%%%%%%%%%%%%%%%%%%%%%%%%%%%
This implies that isospin non-conserving interactions are much weaker 
than the electromagnetic ones which are much weaker than the isospin 
conserving strong ones. 
In fact, assuming that the isospin non-conservation is caused by the 
$\eta\pi^0$ mixing with the mixing parameter, $\epsilon \simeq 10^{-2}$, 
as usual~\cite{Dalitz}, and applying the vector meson dominance 
(VMD)~\cite{VMD} to the radiative decay, we can easily reproduce 
Eq.~(\ref{eq:ratio-D_s^*-exp}), i.e., $R(D_s^{*+}) \simeq 0.06$. 
Next, when $D_{s0}^+(2317)$ is assigned to the iso-triplet tetra-quark 
scalar $\hat F_I^+\sim [cn][\bar s \bar n]_{I=1}$, 
Eq.~(\ref{eq:ratio-D_{s0}}) can be satisfied~\cite{HT-isospin}, i.e., 
%%%%%%%%%%%%%%%%%%%%%%%%%%%%%%%%%%%%%%%%%%%%%%%%%%%%%%%%%%%%%%%%%%%%%%%
$R(D_{s0}^+(2317)=\hat F_I^+)\sim (4-5)\times 10^{-3} \ll 0.059$.  
%%%%%%%%%%%%%%%%%%%%%%%%%%%%%%%%%%%%%%%%%%%%%%%%%%%%%%%%%%%%%%%%%%%%%%%
In contrast, if $D_{s0}^+(2317)$ were assigned to an iso-singlet state, 
(i) the conventional scalar $D_{s0}^{*+}\sim \{c\bar s\}$, or (ii) the 
iso-singlet tetra-quark $\hat F_0^+\sim [cn][\bar s\bar n]_{I=0}$, 
Eq.~(\ref{eq:ratio-D_{s0}}) could not be satisfied, i.e., 
(i) $R(D_{s0}^+(2317)=D_{s0}^{*+})\sim 70 \gg 0.059$, and 
(ii) $R(D_{s0}^+(2317)=\hat F_0^+)\sim 3 \gg 0.059$, as expected above.   
In this way, it is seen that $D_{s0}^+(2317)$ should be assigned to 
an iso-triplet state $\hat F_I^+$. 
In addition, we have learned that $\hat F_0^+$ and $D_{s0}^{*+}$ decay 
dominantly into radiative channels. 
For more details, see Refs.~\cite{HT-isospin} and \cite{ECT-talk}. 

Just after the discovery of $D_{s0}^+(2317)$, charm-strange scalar 
mesons which are degenerate with $D_{s0}^+(2317)$ have been observed not 
only in the $D_s^+\pi^0$ but also the $D_s^{*+}\gamma$ channels of $B$ 
decays~\cite{Belle-D_{s0}},   
%%%%%%%%%%%%%%%%%%%%%%%%%%%%%%%%%%%%%%%%%%%%%%%%%%%%%%%%%%%%%%%%%%%%%%%%
$Br(B \rightarrow \bar{{D}}
\tilde{{D}}_{{s0}}^{{+}}{(2317)}[{D_s}^{{+}}{\pi^0}])
=(8.5^{{+}2.1}_{{-}1.9}{\pm} 2.6)\times 10^{-4}$ and 
${Br}({B}\rightarrow\bar{{D}}\tilde{{D}}_{{s0}}^{{+}}{(2317)}
{[D_s^{{*+}}{\gamma}])}
{=(2.5^{{+}2.0}_{{-}1.8}({<} 7.5))\times 10^{-4}}$. 
%%%%%%%%%%%%%%%%%%%%%%%%%%%%%%%%%%%%%%%%%%%%%%%%%%%%%%%%%%%%%%%%%%%%%%%%
(The above naming conventions, 
%%%%%%%%%%%%%%%%%%%%%%%%%%%%%%%%%%%%%%%%%%%%%%%%%%%%%%%%%%%%%%%%%%%%%%%%
$\tilde{D}_{s0}^{+}(2317)[D_s^{+}\pi^0]$ and 
$\tilde{D}_{s0}^{+}(2317)[D_s^{*+}\gamma]$, 
%%%%%%%%%%%%%%%%%%%%%%%%%%%%%%%%%%%%%%%%%%%%%%%%%%%%%%%%%%%%%%%%%%%%%%%%
have been taken to distinguish the charm-strange scalar mesons observed 
in $B$ decays from  $D_{s0}^+(2317)$ in $e^+e^-$ annihilation.) 
It should be noted that the above production rate of 
$\tilde{D}_{s0}^{+}(2317)[D_s^{*+}\gamma]$ seems to be not much smaller 
than that of $\tilde{D}_{s0}^{+}(2317)[D_s^{+}\pi^0]$, in contrast 
with the $e^+e^-$ annihilation. 
We now identify~\cite{D_{s0}-KT} 
$\tilde{D}_{s0}^{+}(2317)[D_s^{+}\pi^0]$ with $D_{s0}^+(2317)$ which has 
been assigned to $\hat F_I^+$, and 
$\tilde{D}_{s0}^{+}(2317)[D_s^{*+}\gamma]$ is assigned to $\hat F_0^+$ 
which decays dominantly into the $D_s^{*+}\gamma$, as discussed before. 
It should be noted that $\hat F_I^+$ and $\hat F_0^+$ are degenerate 
with each other, in analogy to $a_0(980)$ and $f_0(980)$.  

On the other hand, mass of the charm-strange ($\mathcal{C}=S=1$) scalar 
state has recently been calculated on the lattice~\cite{KF-Liu}, and the 
result has reproduced the measured mass of $D_{s0}^+(2317)$ which has 
been naturally assigned to the iso-triplet $\hat F_I^+$ in the above. 
This implies that the mass of the lowest $\mathcal{C}=S=1$ state which 
can contain not only the scalar $\{c\bar s\}$ but also the scalar 
$[cn][\bar s\bar n]_{I=0}$, etc. is much lower than that of the scalar 
$\{c\bar s\}$ which has been calculated in the quench approximation 
(i.e., with no multi-quark component) on the lattice~\cite{Bali}, and 
hence the lowest $\mathcal{C}=S=1$ state cannot be dominated by the 
$\{c\bar s\}$ but could be by the $[cn][\bar s\bar n]_{I=0}$ component. 
It would be natural because $a_0(980)$ and $f_0(980)$ have been 
assigned~\cite{Jaffe} to the scalar $[ns][\bar n\bar s]_{I=1}$ and 
$[ns][\bar n\bar s]_{I=0}$, and are approximately degenerate with each 
other while $f_0(1500)$ which is expected~\cite{CT} to be dominated by 
the scalar $\{s\bar s\}$ is much heavier. 

Because $D_{s0}^+(2317)$ has been assigned to $\hat F_I^+$, its neutral 
and doubly charged partners, $\hat F_I^0$ and $\hat F_I^{++}$, should 
exist, although they have not been observed in inclusive $e^+e^-$ 
annihilation~\cite{Babar-constraint}. 
This implies that their production is suppressed in this process, 
as was understood within the framework of minimal $q\bar q$ pair 
creation~\cite{production-D_{s0}-KT}. 
In this way, it can be understood why experiments did not observe 
them~\cite{Scadron-70}. 
In addition, it has been 
discussed~\cite{production-D_{s0}-KT,Scadron-70} that 
it is better to search for them in $B$ decays, because the 
$\tilde{D}_{s0}^{+}(2317)[D_s^{*+}\gamma]$ as a signal of $\hat F_0^+$ 
has already been observed  in $B$ decays, as mentioned above, and that 
their production rates are expected to be 
%%%%%%%%%%%%%%%%%%%%%%%%%%%%%%%%%%%%%%%%%%%%%%%%%%%%%%%%%%%%%%%%%%%%%%%%
\begin{eqnarray}
&&
Br(B_u^+\rightarrow D^-\hat F_I^{++})
\sim Br(B_u^+\rightarrow \bar D^0\tilde D_{s0}(2317)[D_s^+\pi^0])_
{\rm exp}                                                   \nonumber\\
&& \sim 
Br(B_d^0\rightarrow \bar D^0\hat F_I^{0})
\sim Br(B_d^0\rightarrow D^-\tilde D_{s0}(2317)[D_s^+\pi^0])_
{\rm exp}\sim 10^{-(3 - 4)},     
\end{eqnarray}
%%%%%%%%%%%%%%%%%%%%%%%%%%%%%%%%%%%%%%%%%%%%%%%%%%%%%%%%%%%%%%%%%%%%%%%%
because all these decays can be described by similar quark-line 
diagrams, where more precise values of their measurements have been 
given in Refs.~\cite{Belle-D_{s0}} and \cite{Babar-B-decay}. 
In addition to $\hat F_I^{0,+,++}$ and $\hat F_0^+$, the 
$[cq][\bar q\bar q]$ states can have a narrow~\cite{D_{s0}-KT} 
$\hat D\sim [cn][\bar u\bar d]$. 
This, as well as the conventional $D_0^*$, should be found in the 
observed $D\pi$ enhancement just below the well-known $D_2^*$ peak. 
Therefore, we now investigate the conventional open-charm scalar 
mesons, $D_0^*$ and $D_{s0}^{*+}$, to distinguish them from tetra-quark 
$\hat D$ and $\hat F_I^+$. 
The most recent measurement of the $D\pi$ enhancement~\cite{Babar-D_0} 
has provided $m_{D_0} = 2297\pm 32$ MeV and $\Gamma({D_0})= 273\pm 74$ 
MeV. 
However, it is expected that the above very broad enhancement might 
have a structure~\cite{McKellar-KT} containing a broad conventional 
scalar $D_0^*$ and a narrow tetra-quark $\hat D$.  
Although the latter seems to have already been observed as a narrow peak 
around the lower tail of the $D\pi$ enhancement, it has not seriously 
been considered in Ref.~\cite{Babar-D_0}. 
Because masses of $D_0^*$ and $D_{s0}^{*+}$ are not definitely known 
yet, as seen above, we tentatively take $m_{D_0^*}\simeq 2.3$ GeV and 
$m_{D_{s0}^{*+}}\simeq 2.4$ GeV. 
The latter seems to be compatible with a prediction on the scalar 
$\{c\bar s\}$ mass in the quench approximation~\cite{Bali}, as 
mentioned before. 
Taking the flavor $SU_f(4)$ relation for the strong vertices with a 
$20-30$ \% deviation of spatial wf. overlap from unity (the symmetry 
limit)~\cite{HT-isospin} and the experimental data~\cite{PDG08} on 
the well-known light scalar $K_0^*$ as the input data, rates for their 
dominant decays, $D_0^*\rightarrow D\pi$ and 
$D_{s0}^{*+}\rightarrow DK$, and hence their widths can be estimated 
to be $\Gamma(D_0^*)\simeq 50 - 60$ MeV and 
$\Gamma(D_{s0}^{*+})\simeq 40 - 50$ MeV~\cite{McKellar-KT}. 
The latter leads to 
$\Gamma(D_{s0}^{*+}\rightarrow D_s^+\pi^0)\simeq 0.2 - 0.3$ keV, 
and hence $R(D_{s0}^{*+})\sim 70$, as discussed before. 
The above $\Gamma(D_0^*)$ is much smaller than the width of the measured 
broad $D\pi$ enhancement mentioned before. 
Therefore, we expect that the observed broad $D\pi$ enhancement can have 
a structure which includes the broad $D_0^*$ and the narrow $\hat D$, as 
discussed before. 
The CDF~\cite{Shapiro} also observed peaks in $D\pi$ mass distributions 
around $2.2 - 2.3$ GeV which can include $\hat D$ and $D_0^*$. 
Besides, a clear peak in $DK$ mass distribution around $2.4$ GeV which is 
degenerate with  $D_{s0}^{*+}$ has been observed by the CLEO~\cite{Kubota}. 
Because these peaks have been taken away as false peaks, however,  
we hope that experiments re-analyze more precisely the above mass 
distributions and find true signals of $D_{s0}^{*+}$, $D_0^*$ and 
$\hat D$ behind the false peaks. 

%%%%%%%%%%%%%%%%%%%%%%%%%%%%%%%%%%%%%%%%%%%%%%%%%%%%%%%%%%%%%%%%%%%%%%%%
\section{Hidden-Charm Mesons}
%%%%%%%%%%%%%%%%%%%%%%%%%%%%%%%%%%%%%%%%%%%%%%%%%%%%%%%%%%%%%%%%%%%%%%%%
$X(3872)$ was discovered in the $\pi^+\pi^-J/\psi$ mass distribution by 
the Belle~\cite{Belle-X(3872)}, and then 
confirmed~\cite{confirmation-of-X} by the CDF, D0 and Babar. 
(Hereafter, we describe $J/\psi$ as $\psi$, for simplicity.) 
Experiments~\cite{Belle-X-J^P,Belle-X-gamma-psi,Babar-X-gamma-psi} favor 
$1^{++}$ as the $J^{P\mathcal{C}}$ of $X(3872)$. 
However, it decays into two different final states with opposite  
$G$-parities, 
%%%%%%%%%%%%%%%%%%%%%%%%%%%%%%%%%%%%%%%%%%%%%%%%%%%%%%%%%%%%%%%%%%%%%%%%
\begin{equation}
R\equiv \frac{Br(X(3872)\rightarrow \pi^+\pi^-\pi^0\psi)}
{Br(X(3872)\rightarrow \pi^+\pi^-\psi)}
=1.0\pm 0.4\pm 0.3.                           \label{eq:3pi/2pi}
\end{equation}
%%%%%%%%%%%%%%%%%%%%%%%%%%%%%%%%%%%%%%%%%%%%%%%%%%%%%%%%%%%%%%%%%%%%%%%%
This is puzzling because the well-known strong interactions conserve 
$G$-parity. 
In addition, the Belle~\cite{Belle-X(3872)} and CDF~\cite{CDF-pipi} 
have noted that the decay $X(3872)\rightarrow \pi^+\pi^-\psi$ 
proceeds through $\rho^0\psi$. 
If the isospin were conserved in the decay, there should exist charged 
partners of $X(3872)$, in contradiction to a negative result from an 
experimental search~\cite{Babar-X-charged-partner}.  
This would imply that $X(3872)$ is an iso-singlet state, and hence the
isospin conservation does not work in the  
$X(3872)\rightarrow\rho^0\psi\rightarrow\pi^+\pi^-\psi$ decay. 
Besides, the Belle~\cite{Belle-X-omega} has suggested  
that the $X(3872)\rightarrow\pi^+\pi^-\pi^0\psi$ decay proceeds 
through the sub-threshold $X(3872)\rightarrow \omega \psi$. 
If isospin is conserved in this decay, $X(3872)$ would be an 
iso-singlet state. 
This is consistent with the above negative result on the search for its 
charged partners. 

Although various approaches~\cite{various-approaches} to solve the above 
puzzle have been proposed, they are unnatural, because the 
phenomenologically well-known $\omega\rho^0$ 
mixing~\cite{PDG08,omega-rho} 
which can play an important role in the isospin non-conservation under 
consideration~\cite{omega-rho-KT} has not been considered. 
%taken into account. 
Under the assumption that the above isospin non-conservation is caused 
by the $\omega\rho^0$ mixing with a mixing parameter\cite{omega-rho-KT} 
$|g_{\omega\rho}|\simeq 3.4\times 10^{-3}$ GeV$^2$,  
%which is consistent with the one in Ref.~\citen{omega-rho}, 
the isospin non-conserving 
$X(3872)\rightarrow \rho^0\psi$ decay proceeds through two steps; 
the isospin conserving %sub-threshold decay, 
$X(3872)\rightarrow \omega\psi$  
and the subsequent $\omega\rho^0$ mixing, 
$X(3872)\rightarrow\omega\psi\rightarrow\rho^0\psi$. 
Here we consider the $X(3872)\rightarrow \gamma\psi$ in place of 
the $X(3872)\rightarrow\pi^+\pi^-\pi^0\psi$ decay in 
Eq.~(\ref{eq:3pi/2pi}), because the kinematics of the former is much 
simpler than the latter. 
As the result, we shall see below that existing data on the ratio 
%%%%%%%%%%%%%%%%%%%%%%%%%%%%%%%%%%%%%%%%%%%%%%%%%%%%%%%%%%%%%%%%%%%%%%%%
\begin{equation}
R^\gamma_X \equiv \frac{Br(X(3872)\rightarrow \gamma\psi)}
     {Br(X(3872)\rightarrow \pi^+\pi^-\psi)}   
                                      \label{eq:radiative-fraction-def}
\end{equation}
%%%%%%%%%%%%%%%%%%%%%%%%%%%%%%%%%%%%%%%%%%%%%%%%%%%%%%%%%%%%%%%%%%%%%%%%
will select a realistic interpretation of $X(3872)$. 
When the above assumption is combined with the VMD~\cite{VMD}, the 
$X(3872)\rightarrow \gamma\psi$ decay would proceed as 
%%%%%%%%%%%%%%%%%%%%%%%%%%%%%%%%%%%%%%%%%%%%%%%%%%%%%%%%%%%%%%%%%%%%%%%%
\begin{eqnarray}
&& X(3872)\rightarrow\omega\psi\rightarrow\gamma\psi
\,\,\,\,{\rm and}\,\,\,\, 
X(3872)\rightarrow\omega\psi\rightarrow\rho^0\psi \rightarrow 
\gamma\psi.                       
                                                \label{eq:X-gamma}
\end{eqnarray}
%%%%%%%%%%%%%%%%%%%%%%%%%%%%%%%%%%%%%%%%%%%%%%%%%%%%%%%%%%%%%%%%%%%%%%%%
However, the contribution of the second decay is much smaller than that 
for the first one because $|g_{\omega\rho}/m_\omega^2| \ll 1$, while the 
role of the $\rho^0$ pole can be strongly enhanced~\cite{omega-rho-KT} 
in the 
%%%%%%%%%%%%%%%%%%%%%%%%%%%%%%%%%%%%%%%%%%%%%%%%%%%%%%%%%%%%%%%%%%%%%%%%
$X(3872)\rightarrow\omega\psi\rightarrow\rho^0\psi\rightarrow 
\pi^+\pi^-\psi$ 
%%%%%%%%%%%%%%%%%%%%%%%%%%%%%%%%%%%%%%%%%%%%%%%%%%%%%%%%%%%%%%%%%%%%%%%%
because  
%%%%%%%%%%%%%%%%%%%%%%%%%%%%%%%%%%%%%%%%%%%%%%%%%%%%%%%%%%%%%%%%%%%%%%%%
$|g_{\omega\rho}/(m_\omega^2 - m_\rho^2)| \gg 
|g_{\omega\rho}/m_\omega^2|$.  
%%%%%%%%%%%%%%%%%%%%%%%%%%%%%%%%%%%%%%%%%%%%%%%%%%%%%%%%%%%%%%%%%%%%%%%%

If $X(3872)$ were an axial-vector charmonium, the radiative 
decay under consideration could have an extra contribution through the 
$\psi$ pole, $X(3872)\rightarrow\psi\psi\rightarrow\gamma\psi$, as the  
dominant one. 
In contrast, when $X(3872)$ is a tetra-quark state 
like~\cite{Terasaki-X} $\{[cn](\bar c\bar n) + (cn)[\bar c\bar n]\}$ 
arising from the last term on the r.h.s. of Eq.~(\ref{eq:4-quark}), 
such a contribution is suppressed because of the OZI rule~\cite{OZI}. 
Therefore, we study if the above isospin non-conservation can be 
reconciled with the measured ratios,
%%%%%%%%%%%%%%%%%%%%%%%%%%%%%%%%%%%%%%%%%%%%%%%%%%%%%%%%%%%%%%%%%%%%%%%%
$(R^\gamma_X)_{\rm Belle} = 0.14 \pm 0.05$~\cite{Belle-X-omega} 
and $(R^\gamma_X)_{\rm Babar} = 0.33 \pm 0.12$~\cite{Babar-gamma-psi'}. 
%%%%%%%%%%%%%%%%%%%%%%%%%%%%%%%%%%%%%%%%%%%%%%%%%%%%%%%%%%%%%%%%%%%%%%%%

In the above $\omega\rho^0$ mixing model~\cite{omega-rho-KT}, the 
value of $R^\gamma_X$ in Eq.~(\ref{eq:radiative-fraction-def}) can be 
estimated without any unknown parameter, if $X(3872)$ is a tetra-quark 
system, i.e., 
%%%%%%%%%%%%%%%%%%%%%%%%%%%%%%%%%%%%%%%%%%%%%%%%%%%%%%%%%%%%%%%%%%%%%%%%
$(R^\gamma_X)_{\rm tetra}\simeq (R^\gamma_X)_{\rm Babar} 
\sim (R^\gamma_X)_{\rm Belle}$,
%%%%%%%%%%%%%%%%%%%%%%%%%%%%%%%%%%%%%%%%%%%%%%%%%%%%%%%%%%%%%%%%%%%%%%%%
because all the parameters involved in the decays can be estimated by 
using the existing experimental data~\cite{PDG08}, except for the  
$X\omega\psi$ coupling $g_{X\omega\psi}$ which can be canceled by taking 
the ratio of decay rates in Eq.~(\ref{eq:radiative-fraction-def}). 
(The $\gamma V$ coupling strengths $X_V(0)$, 
$V=\rho^0,\,\omega,\,\phi,\,\psi$, on the photon-mass-shell have already 
been estimated~\cite{VMD-Terasaki}.) 
In addition, the measured production of \underline{prompt} $X(3872)$ 
seems to favor a more compact object (i.e., a tetra-quark meson) over a 
loosely bound meson-meson molecule~\cite{CDF-prompt-X}. 
In contrast, if $X(3872)$ were a charmonium, the estimated ratio would 
be much larger than the observation, i.e., 
%%%%%%%%%%%%%%%%%%%%%%%%%%%%%%%%%%%%%%%%%%%%%%%%%%%%%%%%%%%%%%%%%%%%%%%%
$(R^\gamma_X)_{c\bar c}\gg (R^\gamma_X)_{\rm tetra}
\simeq (R^\gamma_X)_{\rm Babar}
\sim (R^\gamma_X)_{\rm Belle}$, 
%%%%%%%%%%%%%%%%%%%%%%%%%%%%%%%%%%%%%%%%%%%%%%%%%%%%%%%%%%%%%%%%%%%%%%%%
because of the OZI rule. 
Therefore, the existing data on $R^\gamma_X$ favor a tetra-quark 
interpretation of $X(3872)$, although a small mixing of $\chi'_{c1}$ 
would be needed to understand the measured 
ratio~\cite{Babar-gamma-psi'}, 
%%%%%%%%%%%%%%%%%%%%%%%%%%%%%%%%%%%%%%%%%%%%%%%%%%%%%%%%%%%%%%%%%%%%%%%%
${\Gamma(X\rightarrow\gamma\psi')}/
{\Gamma(X\rightarrow\gamma\psi)}|_{\rm Babar} = 3.4 \pm 1.4$. 
%%%%%%%%%%%%%%%%%%%%%%%%%%%%%%%%%%%%%%%%%%%%%%%%%%%%%%%%%%%%%%%%%%%%%%%%
See Ref.~\cite{omega-rho-KT} for more details. 

%%%%%%%%%%%%%%%%%%%%%%%%%%%%%%%%%%%%%%%%%%%%%%%%%%%%%%%%%%%%%%%%%%%%%%%%
\section{Summary}
%%%%%%%%%%%%%%%%%%%%%%%%%%%%%%%%%%%%%%%%%%%%%%%%%%%%%%%%%%%%%%%%%%%%%%%%

Comparing the ratio of rates for the 
$D_{s0}^{+}(2317)\rightarrow D_s^{*+}\gamma$ decay to the $D_s^+\pi^0$ 
with the experimental constraint Eq.~(\ref{eq:ratio-D_{s0}}), we have 
seen that assigning $D_{s0}^+(2317)$ to $\hat F_I^+$ is favored by 
experiments. 
In this case, $\hat F_I^0$, $\hat F_I^{++}$ and $\hat F_0^+$ should 
exist and be observed. 
However their production through inclusive $e^+e^-\rightarrow c\bar c$ 
is suppressed, so that their observation is likely to be quite 
difficult, although $D_{s0}^+(2317)$ itself has already been observed. 
Therefore, we have discussed that, to search for them, $B$ decays would 
be much better. 
In fact, an indication of 
$\hat F_0^+ = \tilde D_{s0}^+(2317)[D_s^{*+}\gamma]$ 
has already been observed by the Belle~\cite{Belle-D_{s0}}. 

We have studied the ratio of decay rates $R^\gamma_X$ in 
Eq.~(\ref{eq:radiative-fraction-def}), assuming that the isospin 
non-conservation is caused by the phenomenologically well-known 
$\omega\rho^0$ mixing. 
As the result, we have seen that the existing data on $R^\gamma_X$ and 
production of the prompt $X(3872)$ favor a tetra-quark interpretation 
of $X(3872)$ like $\{[cn](\bar c\bar n) + (cn)[\bar c\bar n]\}_{I=0}$ 
over a meson-meson molecule and a charmonium. 
To confirm the above interpretation, observation of 
$\{[cn](\bar c\bar n) - (cn)[\bar c\bar n]\}_{I=0}$ with a mass close 
to $m_{X(3872)}$ in the $\pi^0\pi^0\psi$ 
%%%%%%%%%%%%%%%%%%%%%%%%%%%%%%%%%%%%%%%%%%%%%%%%%%%%%%%%%%%%%%%%%%%%%%%
channel %~\cite{omega-rho-KT} 
%%%%%%%%%%%%%%%%%%%%%%%%%%%%%%%%%%%%%%%%%%%%%%%%%%%%%%%%%%%%%%%%%%%%%%%
is awaited. 

%%%%%%%%%%%%%%%%%%%%%%%%%%%%%%%%%%%%%%%%%%%%%%%%%%%%%%%%%%%%%%%%%%%%%%%
\section*{Acknowledgments} 
The author would like to thank Yukawa Institute for Theoretical 
Physics (YITP) at Kyoto University. 
Discussions during the workshop on {\it New Frontier in QCD 2010} at 
YITP were useful to complete this work. 
He also would like to appreciate the organizers for financial supports. 
%%%%%%%%%%%%%%%%%%%%%%%%%%%%%%%%%%%%%%%%%%%%%%%%%%%%%%%%%%%%%%%%%%%%%%%

%%%%%%%%%%%%%%%%%%%%%%%%%%%%%%%%%

%\end{references}
%%%%%%%%%%%%%%%%%%%%%%%

\begin{thebibliography}{99}
\setlength{\itemsep}{2pt} 

{%\normalsize 

\bibitem{Jaffe} 
R.~L.~Jaffe, Phys. Rev. D {\bf 15} (1977), 267 and 281. 

\bibitem{D_{s0}-KT} 
K.~Terasaki, Phys. Rev. D {\bf 68} (2003), 011501(R). 

\bibitem{color}
M.~Y.~Han and Y.~Nambu, Phys. Rev. {\bf 139} (1965), B 1006;  
S.~Hori, Prog. Theor. Phys. {\bf 36} (1966), 131.

\bibitem{PDG08} 
C.~Amsler et al., the Particle Data Group, Phys. Lett.  
{\bf B667} (2008), 1. % , and references quoted therein. 

\bibitem{K-pi-3/2}
P.~Estabrooks et al., Nucl. Phys. {\bf B133} (1978), 490. 

\bibitem{KT-Hadron2003}
K.~Terasaki, AIP Conf. Proc. {\bf 717} (2004), 556; hep-ph/0309279.   

\bibitem{HT-isospin} 
A.~Hayashigaki and K.~Terasaki, Prog. Theor. Phys. {\bf 114} (2005), 
1191; hep-ph/0410393. 

\bibitem{ECT-talk} 
K.~Terasaki, Invited talk at the workshop on {\it Resonances in QCD}, 
July 11 -- 15, 2005, ECT*, Trento, Italy; hep-ph/0512285.

\bibitem{Babar-e^+e^-D_{s0}}  
B.~Aubert et al., Babar Collaboration, Phys. Rev. Lett. {\bf 90} 
(2003), 242001. 

\bibitem{CLEO-D_{s0}} %Observation of $D_{s0}(2317)$ and $D_{s1}(2460)$
D.~Besson et al., the CLEO Collaboration, Phys. Rev. D {\bf 68} (2003), 
032002. 

\bibitem{Dalitz}
R.~H.~Dalitz and F.~von Hippel, Phys. Lett. {\bf 10} (1964), 153.     

\bibitem{VMD} 
M.~Gell-Mann and F.~Zachariasen, Phys. Rev. {\bf 124} (1961), 953. 
%; J.~J.~Sakurai, {\it Currents and Mesons} (Chicago, Ill., 1969). 

\bibitem{Belle-D_{s0}}
P.~Krokovny et al., Belle Collaboration, Phys. Rev. Lett. {\bf 91} 
(2003), 262002. 

\bibitem{KF-Liu} 
K.-F.~Liu, Invited talk at this workshop. 

\bibitem{Bali} G.~S.~Bali, Phys. Rev. D {\bf 68} (2003), 071501(R). 

\bibitem{CT} %Conventional light scalar
F.~E.~Close and N.~A.~T\"ornquvist, J. Phys. G {\bf 28} (2002), R249. 

\bibitem{Babar-constraint}
B.~Aubert et al., Babar Collaboration, Phys. Rev. D {\bf 74} (2006), 
032007. %; hep-ex/0604031.  

\bibitem{production-D_{s0}-KT}
K.~Terasaki, Prog. Theor. Phys. {\bf 116} (2006), 435; hep-ph/0604207.  

\bibitem{Scadron-70}
K.~Terasaki, AIP Conf. Proc. {\bf 1030} (2008), 190; hep-ph/0804.2295.  

\bibitem{Babar-B-decay}
E.~Robutti, Babar Collaboration, Acta Phys. Polon. {\bf B36} (2005), 
2315. 

\bibitem{Babar-D_0}
B.~Aubert, et al., Babar Collaboration, hep-ex/0901.1291v2. 

\bibitem{McKellar-KT}   %Conventional charmed scalar mesons
K.~Terasaki, hep-ph/0311069; 
K.~Terasaki and Bruce H J McKellar, Prog. Theor. Phys. {\bf 114} 
(2005), 205; hep-ph/0501188.

\bibitem{Kubota} 
Y.~Kubota et al., CLEO Collaboration, Phys. Rev. Lett. 
{\bf 72} (1994), 1972. 

\bibitem{Shapiro} 
M.~Shapiro, CDF Collaboration, Flavor Physics and CP Violation 
(FPCP) conference, Paris, June 3 -- 6, 2003.  

%%%%%%%%%%%%%%%%%%%%%%%%%%%%%%%%%%%%%%%%%%%%%%%%%%%%%%%%%%%%%%%%%%%%%%%
%Observation of $X(3872)$
\bibitem{Belle-X(3872)}
S.~K.~Choi et al., Belle Collaboration, Phys. Rev. Lett. {\bf 91} 
(2003), 262001. 

\bibitem{confirmation-of-X}
%\bibitem{CDF-X(3872)}
D.~Acosta et al., CDF Collaboration, Phys. Rev. Lett. {\bf 93} (2004), 
072001; 
%
%\bibitem{D0-X(3872)}
V.~M.~Abazov et al., D0 Collaboration, Phys. Rev. Lett. {\bf 93} 
(2004), 162002; 
%
%\bibitem{Babar-X(3872)}
B.~Aubert et al., Babar Collaboration, Phys. Rev. D {\bf 71} (2005), 
071103. 
%%%%%%%%%%%%%%%%%%%%%%%%%%%%%%%%%%%%%%%%%%%%%%%%%%%%%%%%%%%%%%%%%%%%%%%

\bibitem{Belle-X-J^P} 
K.~Abe et al., Belle Collaboration, hep-ex/0505038.

\bibitem{Belle-X-gamma-psi} 
K.~Abe et al., Belle Collaboration, hep-ex/0505037. 

\bibitem{Babar-X-gamma-psi} 
B.~Aubert et al., Babar Collaboration, Phys. Rev. D {\bf 74} (2006), 
071101(R); hep-ex/0809.0042.

\bibitem{CDF-pipi} 
A.~Abulencia et al., CDF Collaboration, Phys. Rev. Lett. {\bf 96} 
(2006), 102002.  

\bibitem{Babar-X-charged-partner}
B.~Aubert et al., Babar Collaboration, Phys. Rev. D {\bf 71} (2005), 
031501.

\bibitem{Belle-X-omega} 
K.~Abe et al., Belle Collaboration, hep-ex/0505037. 

\bibitem{various-approaches}
N.~A.~T\"ornqvist, hep-ph/0308277;     %\bibitem{molecule} 
L.~Maiani, F.~Piccinini, A.~D.~Polosa and V.~Riquer, Phys. Rev. D 
{\bf 71} (2005), 014028;               %\bibitem{Maiani}
D.~Gamermann and E.~Oset, Phys. Rev. D {\bf 80} (2009), 014003; 
%; hep-ph/0905.0402. %\bibitem{Oset2}
M.~Takizawa and S.~Takeuchi, to appear in EPJ web of Conference, 
Few-Body 19.                            %\bibitem{Takizawa} 

\bibitem{omega-rho} 
%C.~McNeile, C.~Michael and C.~Urbach, hep-lat/0902.3897 and references 
%quoted therein. %; 
G.~A.~Miller, A.~K.~Opper and E.~J.~Stephenson, Ann. Rev. Nucl. Part. 
Sci. {\bf 56}, 253 (2006). 

\bibitem{omega-rho-KT}
K.~Terasaki, Prog. Theor. Phys. {\bf 122} (2009), 1285; 
hep-ph/0904.3368v2. 

\bibitem{Terasaki-X}  %A tetra-quark picture of X(3872)
K.~Terasaki, Prog. Theor. Phys. {\bf 118} (2007), 821; 
hep-ph/0706.3944.

\bibitem{OZI}
S.~Okubo, Phys. Lett. {\bf 5} (1963),165; G.~Zweig, CERN Report
No. TH401 (1964); J.~Iizuka, K.~Okada and O.~Shito,
Prog. Theor. Phys. {\bf 35} (1965), 1061. 

\bibitem{Babar-gamma-psi'}
B.~Aubert et al., Babar Collaboration, hep-ex/0809.0042.

\bibitem{VMD-Terasaki} 
K.~Terasaki, Lett. Nuovo Cim. {\bf 31}, 457 (1981); Nuovo Cim. 
{\bf 66A}, 475 (1981). 

\bibitem{CDF-prompt-X}
C.~Bignamini, B.~Grinstein, F.~Piccinini,A.~D.~Polosa and C.~Saballi, 
hep-ph/0906.0882; A.Abulencia et al., CDF Collaboration, Phys. Rev. 
Lett. {\bf 98} (2007), 132002; CDF note 7159 (2004); 
URL http://www-cdf.fnal.gov. 

}
%%%%%%%%%%%%%%%%%%%%%%%%%%%%%%%%%%%%%
\end{thebibliography}
\end{document}